# Effect of Density of state on Isotope Effect Exponent of Two-band Superconductors

P.Udomsamuthirun[(1)], C.Kumvongsa[(2)], A.Burakorn[(1)], P.Changkanarth[(1)] and S.Yoksan[(1)]

[(1)] Department of Physics, Faculty of Science, Srinakharinwirot University, Bangkok 10110, Thailand. E-mail: udomsamut55@yahoo.com

[(2)] Department of Basic Science, School of Science, The University of the Thai Chamber of Commerce, Dindaeng, Bangkok 10400, Thailand.

**Keywords:** isotope effect exponent, two-band superconductors, Density of state

**PACS** : 74.20.Fg, 74.70.Ad.

**Abstract**

The exact formula of $T_c$'s equation and the isotope effect exponent of two-band s-wave superconductors in weak-coupling limit are derived by considering the influence of two kinds of density of state : constant and van Hove singularity. The paring interaction in each band consisted of 2 parts : the electron-phonon interaction and non-electron-phonon interaction are included in our model. We find that the interband interaction of electron-phonon show more effect on isotope exponent than the intraband interaction and the isotope effect exponent with constant density of state can fit to experimental data, $MgB_2$ and high-$T_c$ superconductor, better than van Hove singularity density of state.

## 1. Introduction

The isotope effect exponent, $\alpha$, is one of the most interesting properties of superconductors. In the conventional BCS theory $\alpha = 0.5$ for all element. In high-$T_c$ superconductors, experimenter found that $\alpha$ is smaller than 0.5[1-3]. This unusual small value leads to suggestion that the pairing interaction might be predominantly of electronic origin with a possible small phononnic contribution[4]. To explain the unusual isotope effect in high-$T_c$ superconductors, many models have been proposed such as the van Hove singularity[5-7], anharmonic phonon[8,9], pairing-breaking effect[10], and pseudogap[11,12].

The discovery of [13] of superconductivity in $MgB_2$ with a high critical temperature, $T_c \approx 39$ K, has attracted a lot of considerable attention. Various experiments [14-23] suggest the existence of two-band in $MgB_2$ superconductors. Budko et al.[24] and Hinks et al.[25] measured the isotope effect exponent and estimated as $\alpha \approx 0.3$. The boron isotope exponent is closed to that obtained for the $YNi_2B_2C$ and $LuNi_2B_2C$ borocarbideds [26,27] where theoretical work[28] suggested that the phonons responsible for the superconductivity are high-frequency boron optical modes. This observation is consistent with a phonon-mediated BCS superconducting mechanism that boron phonon modes are playing an important role.

In this paper we start from two-band model that was first introduced by Suhl [29] and Moskalenke[30]. The effect of density of state, van Hove singularity and constant density of state, on the isotope effect exponent of two-band superconductors are investigated in weak-coupling limit. The paring interaction in each band consisted of 2 parts: an attractive electron-phonon interaction and an attractive non-electron-phonon interaction are included in our model. The analytical formula of the isotope effect exponent are derived and computed numerically to compare with experimental data of high-$T_c$ and $MgB_2$ superconductor.

## 2. Model and calculation

The Hamiltonian of the corresponding system is taken in the form

$$H = H_1 + H_2 + H_{12} \tag{1}$$

where $H_1$, $H_2$ and $H_{12}$ are the Hamiltonian of $1^{st}$ band, $2^{nd}$ band and interband respectively that

$$H_1 = \sum_{k\sigma} \varepsilon_{k\sigma} C^+_{k\sigma} C_{k\sigma} - \sum_{kk'} V_{11kk'} C^+_{k\uparrow} C^+_{-k\downarrow} C_{-k'\downarrow} C_{k'\uparrow} \tag{2.1}$$

$$H_2 = \sum_{k\sigma} \varepsilon_{k\sigma} p^+_{k\sigma} p_{k\sigma} - \sum_{kk'} V_{22kk'} p^+_{k\uparrow} p^+_{-k\downarrow} p_{-k'\downarrow} p_{k'\uparrow} \tag{2.2}$$

$$H_{12} = -\sum_{kk'} V_{12kk'} (p^+_{k\uparrow} p^+_{-k\downarrow} C_{-k'\downarrow} C_{k'\uparrow} + C^+_{k\uparrow} C^+_{-k\downarrow} p_{-k'\downarrow} p_{k'\uparrow}) \tag{2.3}$$

Here we use the standard meaning of parameters. $V_{11kk'}$, $V_{22kk'}$, $V_{12kk'}$ are the attractive interaction potential in $1^{st}$ band, $2^{nd}$ band and interband respectively and $V_{12} = V^*_{21}$.

By performing a BCS mean field analysis of Eq.(1) and applying standard techniques, we obtain the gap equation as

$$\Delta_{1k} = -\sum_{k'} V_{11kk'} \frac{\Delta_{1k'}}{2\sqrt{\varepsilon^2_{1k'} + \Delta^2_{1k'}}} \tanh(\frac{\sqrt{\varepsilon^2_{1k'} + \Delta^2_{1k'}}}{2T}) - \sum_{k'} V_{12kk'} \frac{\Delta_{2k'}}{2\sqrt{\varepsilon^2_{2k'} + \Delta^2_{2k'}}} \tanh(\frac{\sqrt{\varepsilon^2_{2k'} + \Delta^2_{2k'}}}{2T})$$

$$\tag{3.1}$$

$$\Delta_{2k} = -\sum_{k'} V_{22kk'} \frac{\Delta_{2k'}}{2\sqrt{\varepsilon_{2k'}^2 + \Delta_{2k'}^2}} \tanh(\frac{\sqrt{\varepsilon_{2k'}^2 + \Delta_{2k'}^2}}{2T}) - \sum_{k'} V_{12kk'} \frac{\Delta_{1k'}}{2\sqrt{\varepsilon_{1k'}^2 + \Delta_{1k'}^2}} \tanh(\frac{\sqrt{\varepsilon_{1k'}^2 + \Delta_{1k'}^2}}{2T})$$

(3.2)

where $\Delta_{1k}$ and $\Delta_{2k}$ are the superconducting order parameters of $1^{st}$ band and $2^{nd}$ band that dependent on wave vector, $\vec{k}$.

Each paring interaction potential consists of 2 parts[31,32] : an attractive electron-phonon interaction $V_{ph}$ and an attractive non-electron-phonon interaction $U_C$. $\omega_D$ and $\omega_c$ is the characteristic energy cutoff of the Debye phonon and non-phonon respectively. The interaction potential $V_{kk'}$ may be written as

$$V_{kk'} = -V_{ph} - U_C \quad \text{for} \quad 0 < |\varepsilon| < \omega_D$$
$$= -U_C \quad \text{for} \quad \omega_D < |\varepsilon| < \omega_c \quad (4)$$

For such as the interaction, the superconducting order parameters can be written as

$$\Delta_{jk} = \Delta_{j1} \quad \text{for} \quad 0 < |\varepsilon| < \omega_D$$
$$= \Delta_{j2} \quad \text{for} \quad \omega_D < |\varepsilon| < \omega_c \quad (5)$$

that $j = 1, 2$. Here $\Delta_{jk}$ represents $\Delta_{1k}$ and $\Delta_{2k}$. $\Delta_{j1}$, $\Delta_{j2}$ are the phonon and non-phonon parts of the order parameters $\Delta_{jk}$ and they are not dependent on wave vector, $\vec{k}$.

**2.1 Constant density of state**

We first consider the two-band superconductors with constant density of state, $N(\varepsilon) = N(0)$, and substitution Eqs.(4) and (5) into Eq.(3). After making some algebra at $T = T_c$, we can get

$$\begin{pmatrix} \Delta_{11} \\ \Delta_{21} \\ \Delta_{12} \\ \Delta_{22} \end{pmatrix} = \begin{pmatrix} (\lambda_1 + \mu_1)I_1 & (\lambda_{12} + \mu_{12})I_1 & \mu_1 I_2 & \mu_{12} I_2 \\ (\lambda_{12} + \mu_{12})I_1 & (\lambda_2 + \mu_2)I_1 & \mu_{12} I_2 & \mu_2 I_2 \\ \mu_1 I_1 & \mu_{12} I_1 & \mu_1 I_2 & \mu_{12} I_2 \\ \mu_{12} I_1 & \mu_2 I_1 & \mu_{12} I_2 & \mu_2 I_2 \end{pmatrix} \begin{pmatrix} \Delta_{11} \\ \Delta_{21} \\ \Delta_{12} \\ \Delta_{22} \end{pmatrix} \quad (6)$$

Here $$I_1 = \int_0^{\omega_D} d\varepsilon \frac{\tanh(\varepsilon/2T_c)}{\varepsilon} \quad (7.1)$$

$$I_2 = \int_{\omega_D}^{\omega_c} d\varepsilon \frac{\tanh(\varepsilon/2T_c)}{\varepsilon} \quad (7.2)$$

The coupling constants are defined as

$$\lambda_1 = N(0)V_{ph}^1, \lambda_2 = N(0)V_{ph}^2, \lambda_{12} = N(0)V_{ph}^{12} = N(0)V_{ph}^{21}$$

and $$\mu_1 = N(0)U_C^1, \mu_2 = N(0)U_C^2, \mu_{12} = N(0)U_C^{12} = N(0)U_C^{21} \quad (8)$$

Solving the secular equation, $I_1$ is given as

$$I_1 = ( \left[ \frac{1}{\mu_t} + \frac{1}{\lambda_t} + I_2\left(\frac{2b_\mu}{\lambda_t} - 1\right) - b_\mu I_2^2 \right] - \left[ \left(\frac{1}{\mu_t} + \frac{1}{\lambda_t} + I_2\left(\frac{2b_\mu}{\lambda_t} - 1\right) - b_\mu I_2^2\right)^2 \right.$$

$$\left. + 4\left(\frac{a_\lambda}{\mu_t} + \frac{b_\mu}{\lambda_t} + \frac{2\lambda_{12}\mu_{12}}{\lambda_t \mu_t} - \frac{(\lambda_2\mu_1 + \lambda_1\mu_2)}{\lambda_t \mu_t} - I_2(a_\lambda + b_\mu) - I_2^2 a_\lambda b_\mu\right)\left(\frac{1}{\lambda_t \mu_t} - \frac{I_2}{\lambda_t} - \frac{b_\mu I_2^2}{\lambda_t}\right) \right]^{1/2} )/$$

$$(-2\left[\frac{a_\lambda}{\mu_t} + \frac{b_\mu}{\lambda_t} + \frac{2\lambda_{12}\mu_{12}}{\lambda_t \mu_t} - \frac{(\lambda_2\mu_1 + \lambda_1\mu_2)}{\lambda_t \mu_t} - I_2(a_\lambda + b_\mu) - I_2^2 a_\lambda b_\mu\right])$$

(9)

where $a_\lambda = \dfrac{\lambda_{12}^2 - \lambda_1\lambda_2}{\lambda_1 + \lambda_2} = \dfrac{\lambda_{12}^2 - \lambda_1\lambda_2}{\lambda_t}$ and $b_\mu = \dfrac{\mu_{12}^2 - \mu_1\mu_2}{\mu_1 + \mu_2} = \dfrac{\mu_{12}^2 - \mu_1\mu_2}{\mu_t}$ .

In case $\omega_D > T_c$, we can make the approximation that $I_1 \approx \ln(\dfrac{1.14\omega_D}{T_c})$ and $I_2 \approx \ln(\dfrac{\omega_C}{\omega_D})$ . The $T_c$'s equation is

$$T_c = 1.14\omega_D \exp( \left[ \frac{1}{\mu_t} + \frac{1}{\lambda_t} + I_2\left(\frac{2b_\mu}{\lambda_t} - 1\right) - b_\mu I_2^2 \right] - \left[ \left(\frac{1}{\mu_t} + \frac{1}{\lambda_t} + I_2\left(\frac{2b_\mu}{\lambda_t} - 1\right) - b_\mu I_2^2\right)^2 \right.$$

$$\left. + 4\left(\frac{a_\lambda}{\mu_t} + \frac{b_\mu}{\lambda_t} + \frac{2\lambda_{12}\mu_{12}}{\lambda_t \mu_t} - \frac{(\lambda_2\mu_1 + \lambda_1\mu_2)}{\lambda_t \mu_t} - I_2(a_\lambda + b_\mu) - I_2^2 a_\lambda b_\mu\right)\left(\frac{1}{\lambda_t \mu_t} - \frac{I_2}{\lambda_t} - \frac{b_\mu I_2^2}{\lambda_t}\right) \right]^{1/2} )/$$

$$(2\left[\frac{a_\lambda}{\mu_t} + \frac{b_\mu}{\lambda_t} + \frac{2\lambda_{12}\mu_{12}}{\lambda_t \mu_t} - \frac{(\lambda_2\mu_1 + \lambda_1\mu_2)}{\lambda_t \mu_t} - I_2(a_\lambda + b_\mu) - I_2^2 a_\lambda b_\mu\right])$$

(10)

Eq.(10) is the $T_c$'s equation of two-band s-wave superconductors with constant density of state that all of the right-hand-side parameters are not depended on $T_c$. Taking an approximation, Eq.(10) become

$$T_c \approx 1.14\omega_D \exp\left(-\left(\frac{1}{\lambda_t \mu_t} - \frac{I_2}{\lambda_t} - \frac{b_\mu I_2^2}{\lambda_t}\right) \middle/ \left(\frac{1}{\mu_t} + \frac{1}{\lambda_t} + I_2\left(\frac{2b_\mu}{\lambda_t} - 1\right) - b_\mu I_2^2\right)\right)$$

$$\approx 1.14\omega_D \exp\left(\frac{(-1/\lambda_t)}{1 + \left(\dfrac{\mu_t}{\lambda_t}\right)\dfrac{1 + 2b_\mu I_2}{1 - \mu_t I_2 - \mu_t b_\mu I_2^2}}\right) \quad (11)$$

The isotope effect exponent will be solved in the harmonic approximation, $\omega_D \propto M^{-1/2}$, and $\omega_c$ does not depend on mass. The isotope effect exponent is defined as

$$\alpha = -\frac{d\ln T_c}{d\ln M}$$

$$= \frac{1}{2} \frac{\omega_D}{T_c} \frac{dT_c}{d\omega_D} \quad (12)$$

where M is the mass of the atom constituting the specimen under consideration.

Substitution Eq.(9) into Eq.(12), we get

$$\alpha_0 = \left(\frac{1}{2}\right)\left(1/(1 - \frac{[B]\tanh[\omega_c/2T_c]}{[A]\tanh[\omega_D/2T_c]})\right) \quad (13)$$

where

$$[A] = \lambda_t + I_1[-2\lambda_t a_\lambda + 4\lambda_{12}\mu_{12} + (\mu_1 - \mu_2)(\lambda_1 - \lambda_2) + I_1\lambda_t\mu_t(b_\mu - a_\lambda)] - I_2\lambda_t\mu_t(1 + 2a_\lambda I_1(1 - I_1 b_\mu))$$

$$- \mu_t\lambda_t b_\mu I_2^2(1 + 2a_\lambda I_1)] \quad (14.1)$$

$$[B] = \mu_t - I_1\{\mu_t(\lambda_t - 2b_\mu) + I_1\lambda_t\mu_t(a_\lambda + b_\mu)\} + 2b_\mu\mu_t(1 - \lambda_t I_1(1 + a_\lambda I_1)) \quad (14.2)$$

Eq.(13) can be easily reduced to isotope effect exponent of BCS theory ($\alpha = 0.5$).

**2.2 The van Hove singularity density of state**

Another case of consideration, we consider the two-band superconductors with the van Hove singularity density of state as

$$N(\varepsilon) = N(0)\ln\left(\frac{E_F}{\varepsilon - E_F}\right)$$

here $E_F$ is Fermi energy. Substitution the van Hove singularity density of state into Eq.(3) and setting new parameters as

$$I_3 = \int_0^{\omega_D} \ln\left|\frac{E_F}{\varepsilon}\right| \frac{\tanh\left(\frac{\varepsilon}{2T_c}\right)}{\varepsilon} d\varepsilon \quad (15.1)$$

and

$$I_4 = \int_{\omega_D}^{\omega_c} \ln\left|\frac{E_F}{\varepsilon}\right| \frac{\tanh\left(\frac{\varepsilon}{2T_c}\right)}{\varepsilon} d\varepsilon \quad (15.2).$$

We can get equation that look like Eq.(6) except we must replace $I_1, I_2$ by $I_3, I_4$ respectively. After some calculation, we can get the equation of $I_3$ that look like Eq.(9). Using an approximation, $I_3 \approx \frac{1}{2}[(\ln(\frac{eE_F}{2T_c}))^2 - (\ln(\frac{E_F}{\omega_D}))^2 + 1]$ and $I_4 \approx \frac{1}{2}\ln(\frac{E_F^2}{\omega_D\omega_c})\ln(\frac{\omega_c}{\omega_D})$, $T_c$ 's equation for van Hove singularity is found as

$$T_c = 1.36 E_F \exp\{-[2(\left[\frac{1}{\mu_t} + \frac{1}{\lambda_t} + I_4\left(\frac{2b_\mu}{\lambda_t} - 1\right) - b_\mu I_4^2\right] - [\left(\frac{1}{\mu_t} + \frac{1}{\lambda_t} + I_4\left(\frac{2b_\mu}{\lambda_t} - 1\right) - b_\mu I_4^2\right)^2$$

$$+ 4\left(\frac{a_\lambda}{\mu_t} + \frac{b_\mu}{\lambda_t} + \frac{2\lambda_{12}\mu_{12}}{\lambda_t \mu_t} - \frac{(\lambda_2 \mu_1 + \lambda_1 \mu_2)}{\lambda_t \mu_t} - I_4(a_\lambda + b_\mu) - I_4^2 a_\lambda b_\mu \right)\left(\frac{1}{\lambda_t \mu_t} - \frac{I_4}{\lambda_t} - \frac{b_\mu I_4^2}{\lambda_t}\right)\right]^{1/2} )/$$

$$(2\left[\frac{a_\lambda}{\mu_t} + \frac{b_\mu}{\lambda_t} + \frac{2\lambda_{12}\mu_{12}}{\lambda_t \mu_t} - \frac{(\lambda_2 \mu_1 + \lambda_1 \mu_2)}{\lambda_t \mu_t} - I_4(a_\lambda + b_\mu) - I_4^2 a_\lambda b_\mu\right]) + \ln^2(\frac{E_F}{\omega_D}) - 1]^{1/2}\}$$

(16.1)

$$\approx 1.36 E_F \exp\{-[(\frac{(2/\lambda_t)}{1 + (\frac{\mu_t}{\lambda_t})\frac{1 + 2b_\mu I_4}{1 - \mu_t I_4 - \mu_t b_\mu I_4^2}}) + \ln^2(\frac{E_F}{\omega_D}) - 1]^{1/2}\}$$

(16.2)

The isotope effect exponent is

$$\alpha_{vhs} = \left(\frac{1}{2}\right)\left(\frac{\ln[E_F/\omega_D]\tanh(\omega_D/2T_c)}{(F(\omega_D) - F(\omega_C)\frac{[B]}{[A]})}\right) \quad (17)$$

Here [A] and [B] are defined as Eq.(14.1) and (14.2), and

$$F(y) = \int_0^{y/2T_c} \ln(\frac{E_F}{2xT_c})\sech^2(x)dx \quad (18)$$

We use Eq.(10) and Eq.(13) to compute numerically and show graph of the isotope exponent with constant density of state versus the critical temperature in Figure.(1). Depending on the measured Debye frequency of $MgB_2$, $\omega_D$ =64.3 meV [24,33] and $T_c \approx 40$ K, the parameters we use for numerical calculation are $\omega_D = 745$ K, $\omega_c = 800$ K. The isotope effect exponent is tend to 0.5 at low $T_c$. However, it can fit well with the experimental data of $MgB_2$ [24] and $(Y_{1-x-y}Pr_xCa_y)Ba_2Cu_3O_{7-y}$ and $YBa_2(Cu_{1-z}Zn)_3O_{7-y}$ [34].

In Figure(2), we use Eq.(16.1) and Eq.(17) to compute numerically and show graph of the isotope exponent with van Hove singularity density of state versus the critical temperature. The parameters are $E_F = 5000$ K, $\omega_D = 745$ K, $\omega_c = 800$ K and the same set of experimental data. We find that this isotope effect exponent can fit well with high-$T_c$ superconductors[34] but it is not for $MgB_2$.

In both figure, the curves that intraband parameters are constants and the interband parameter, $\lambda_{12}$, varied are show more difference of isotope exponent than the curves that interband parameter, $\lambda_{12}$, is constant and the intraband parameters varied with the same ratio. We can conclude that the interband interaction of electron-phonon show more effect on isotope exponent than the intraband interaction. The isotope effect exponent increase as the interband interaction of electron-phonon interaction increase for constant density of state but it is not certainly for van Hove singularity density of state.

## 5. Conclusions

The exact formula of $T_c$'s equation and the isotope effect exponent of two-band s-wave superconductors in weak-coupling limit are derived by considering the influence of two kind of density of state : constant and van Hove singularity . The interband interaction of electron-phonon interaction show more effect on isotope effect exponent than the intraband interaction. The isotope effect exponent with constant density of state can fit to experimental data better than with van Hove singularity density of state and $MgB_2$ can be described by the isotope effect exponent with constant density of state.

**Acknowledgement** The author would like to thank Thailand Research Fund for financial support and, Srinakharinwirot university and the university of the Thai Chamber of Commerce for partial financial support .

**Figure Caption**

**Figure(1).** The graph of the isotope exponent with constant density of state versus $T_c$ is shown. The parameters are $\omega_D = 745$ K, $\omega_C = 800$ K and the experimental data of $MgB_2$ [24] ,and $(Y_{1-x-y} Pr_x Ca_y)Ba_2Cu_3 O_{7-y}$ and $Y Ba_2(Cu_{1-z} Zn)_3 O_{7-y}$ [34].

**Figure.(2).** The graph of the isotope exponent with van Hove singularity density of state versus $T_c$ is shown. The parameters are $E_F = 5000$ K, $\omega_D = 745$ K, $\omega_C = 800$ K and the experimental data of $MgB_2$ [24], and $(Y_{1-x-y} Pr_x Ca_y)Ba_2Cu_3 O_{7-y}$ and $Y Ba_2(Cu_{1-z} Zn)_3 O_{7-y}$ [34].
.

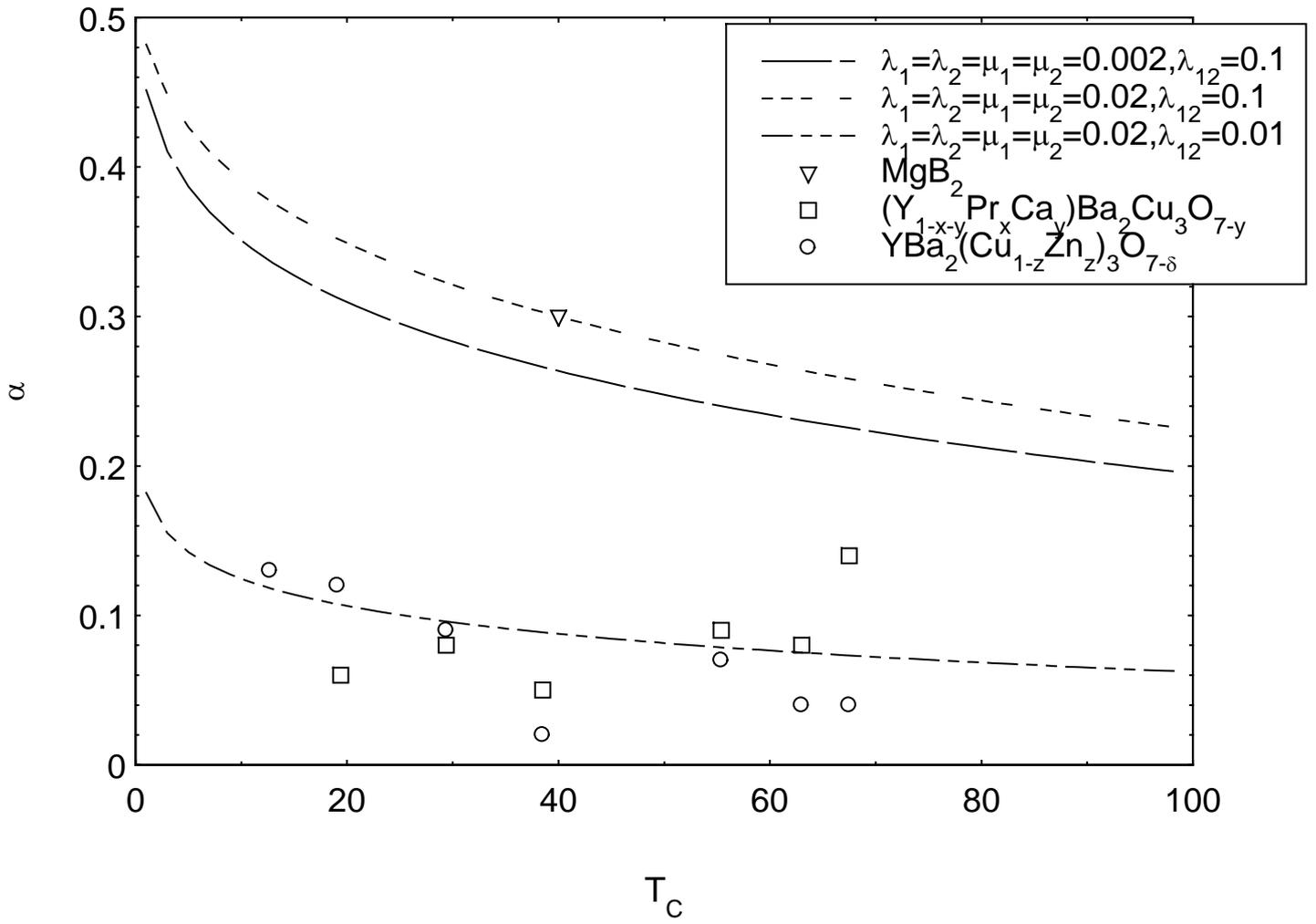

**Figure(1).**

Udomsamuthirun et al.

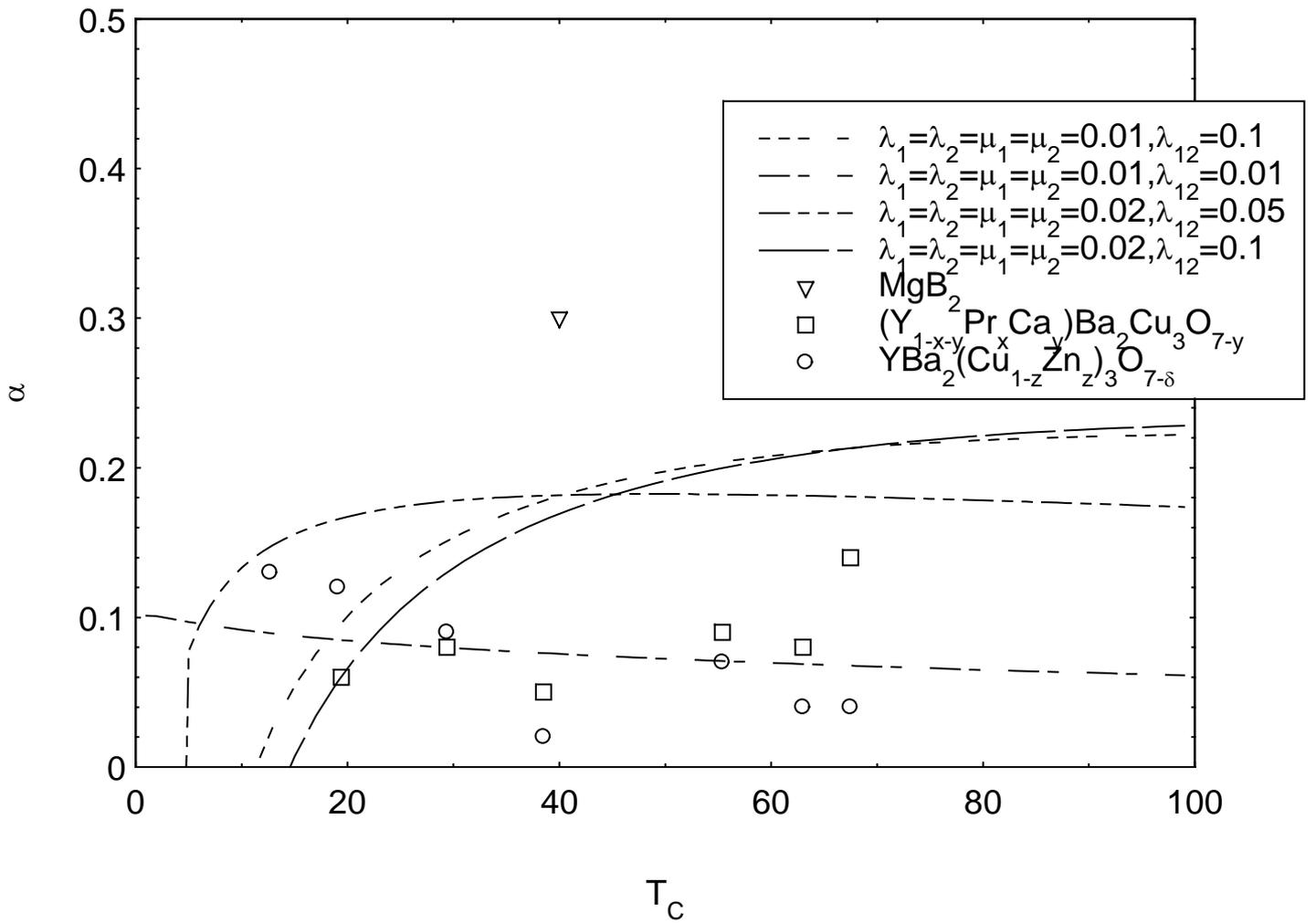

**Figure(2).**

Udomsamuthirun et al.